\documentclass[pre,twocolumn,showpacs,preprintnumbers,amsmath,amssymb]{revtex4}
\usepackage{graphics,epsf}

\begin{document}

\title{Embedding methods for large-scale surface calculations}

\author{J. R.  Trail}\email{j.r.trail@bath.ac.uk}
\author{D. M.  Bird }
\affiliation{ Department of Physics, University of Bath, Bath, UK}

\date{October, 1999}

\begin{abstract}
One of the goals in the development of large scale electronic structure methods is to perform calculations explicitly for a localised region of a system, while still taking into account the rest of the system outside of this region.
An example of this in surface physics would be to embed an adsorbate and a few surface atoms into an extended substrate, hence considerably reducing computational costs.
Here we apply the constrained electron density method of embedding a Kohn-Sham system in a substrate system (first described by P. Cortona\cite{1} and T.A. Wesolowski\cite{2}), within a plane-wave basis and pseudopotential framework.
This approach divides the charge density of the system into substrate and embedded charge densities, the sum of which is the charge density of the actual system of interest.
Two test cases are considered.
First we construct fcc bulk aluminium by embedding one cubic lattice of atoms within another.
Second, we examine a model surface/adsorbate system of aluminium on aluminium and compare with full Kohn-Sham results.
\end{abstract}


\maketitle

\section{DFT Embedding}
Density Functional Theory (DFT) is one of the most powerful tools for the \emph{ab initio} calculation of the physical and chemical properties of materials, being both efficient and accurate.
Many implementations of DFT exist, which differ in the approach taken to approximating the unknown density functional that describes the contribution to the energy of the electrons that is \emph{not} due to the external potential.
The  basic problem of DFT is to minimise the functional
\begin{equation}
E[\rho]=T_s[\rho]+
J[\rho]+E_{xc}[\rho]+\int V_{ext}({\mathbf{r}}) \rho({\mathbf{r}}) d^3{\mathbf{r}}
\end{equation}
where $T_s[\rho]$ is the non-interacting kinetic energy functional, $J[\rho]$ is the Hartree energy, $E_{xc}[\rho]$ is the `exchange-correlation' energy which takes into all non-classical electron-electron interactions, and $V_{ext}({\mathbf{r}})$ is the external potential of the system of interest.
To obtain the ground state energy this must be minimised with respect to variations in $\rho({\mathbf r})$, subject a constrained number of electrons. 
The Hartree and  external potential energies can be simply evaluated for a trial charge density, and accurate approximations are available for $E_{xc}[\rho]$, but no explicit form is known for the non-interacting kinetic energy.
The charge density for the interacting electrons can be identified with the charge density of a `reference' non-interacting electron gas, and this reference system solved self-consistently to yield the energy and charge density for the interacting system - this is the Kohn-Sham method.
Due to orthogonalisation requirements, solving for this reference system scales as $O(N^3)$ where $N$ is the number of electrons which limits the size of system that can be considered\cite{3}.

There are $O(N)$ methods available that take advantage of the `nearsightedness' of the density matrix, but at present these are only applicable where the number of basis functions per atom is small, such as tight-binding or atomic-orbital calculations.
Another approach is to employ approximate kinetic energy functionals, and minimise the functional directly.
Calculations of this sort are cheap and quick, but the approximations available in the literature are generally not sufficient for structural optimisation, let alone chemical accuracy.

Another approach is embedding.
In many cases (eg an adsorbate on a surface) we can divide the system into two regions of space, region $I$ (eg the adsorbate and a few surface atoms) and region $II$ (eg the rest of the surface).
Region $II$ is largely the same as that for a more simple system that may easily be solved for, whereas region $I$ is where all of the interesting physics occurs.
It would obviously be computationally advantageous to solve for region $II$ first, and then solve for region $I$ taking into account the influence of region $II$ in some way.
This `embedding' approach has received a great deal of attention, and a large number of methods have been presented in the literature\cite{4,5,6}.
Many methods approach this as a boundary value problem, at the wavefunction level, but we examine a different approach which starts at the more flexible DFT level, and was first presented by Cortona\cite{1} and Wesolowski\cite{2}.

We start with Eq. (1) written as
\begin{eqnarray}
E[\rho] & = & T_s[\rho_1]+T^{nadd}_s[\rho_1,\rho_2]+T_s[\rho_2] \nonumber \\
        & + & J[\rho]+E_{xc}[\rho]+\int V_{ext}({\mathbf{r}}) \rho({\mathbf{r}}) d^3{\mathbf{r}}
\end{eqnarray}
where $\rho({\mathbf r})=\rho_1({\mathbf r})+\rho_2({\mathbf r})$, $\rho_1({\mathbf r})$ is the charge density of the embedded system that we shall be varying, and $\rho_2({\mathbf r})$ is the substrate charge density that is kept constant.
This defines the non-additive kinetic energy, $T^{nadd}_s[\rho_1,\rho_2]$, as
\begin{equation}
T^{nadd}_s[\rho_1,\rho_2]=T_s[\rho_1+\rho_2]-T_s[\rho_1]-T_s[\rho_2].
\end{equation}

Minimising Eq. (2) with respect to variations in $\rho_1({\mathbf r})$ leads to the Euler-Lagrange equation
\begin{equation}
\frac{\delta T_s[\rho_1]}{\delta \rho_1} +
\frac{\delta T^{nadd}_s[\rho_1,\rho_2]}{\delta \rho_1}
+V_{KS}[\rho;{\mathbf{r}}]=\mu
\end{equation}
where $\mu$ is the chemical potential, $V_{KS}[\rho;{\mathbf{r}}]$ is the usual Kohn-Sham potential associated with density $\rho({\mathbf{r}})$, and a new `embedding potential' term is present.
In the same manner as for the Kohn-Sham case, this leads to the $\rho_1({\mathbf{r}})$ being the solution of the `Kohn-Sham' equations associated with Eq. (4) at self consistency, but with an effective potential given by
\begin{eqnarray}
\lefteqn{
V^{eff}[\rho;{\mathbf{r}}]  =  V_{KS}[\rho;{\mathbf{r}}] +
\frac{\delta T^{nadd}_s[\rho_1,\rho_2]}{\delta \rho_1} } \nonumber \\
&  =  V_{KS}[\rho;{\mathbf{r}}] +
\left(
\frac{\delta T_s[\rho_1+\rho_2]}{\delta \rho_1} - \frac{\delta T_s[\rho_1]}{\delta \rho_1}
\right).
\end{eqnarray}

As applied previously, this method has employed basis function localised to their host atoms, which essentially constrains the charge density to a localised region and to a particular functional form.
These previous applications have been to weakly interacting and insulating systems, whereas here we explore the possibility of examining strongly interacting metallic systems using an unbiased plane wave basis.

\section{Approximate Kinetic Energy Functionals}
As given above we have only re-expressed the original problem in a slightly different form.
In order for this approach to be useful, it must be possible to express the non-additive kinetic-energy accurately.
Of course we do not know an analytic form for this interaction energy, but it is expected to be small, and zero for no overlap between the embedded and substrate systems.
This suggests the use of the approximate kinetic energy functionals available in the literature to construct $T_s^{nadd}$ and its functional derivative.

Two forms of approximate kinetic energy functional are considered.
First a semi-local form, incorporating an `enhancement factor' analogous to the Generalised Gradient Approximation (GGA) of exchange-correlation functionals.
This takes the form
\begin{equation}
T^{enh}_s[\rho]=\frac{3}{5}(3\pi^2)^\frac{2}{3} \int
\rho^\frac{5}{3} F(t) d^3{\mathbf{r}}
\end{equation}
where
\begin{equation}
t=\frac{|\nabla \rho|^2}{\rho^\frac{8}{3}},
\end{equation}
and  $F(t)$ is the enhancement factor.
The Thomas-Fermi functional is a special case of this, where $F(t)=1$, the 1$^{st}$ order gradient expansion corresponds to $F(t)=1+at$ with $a$ an appropriate constant, and the von Weizacker approximation corresponds to $F(t)=bt$ with $b$ again an appropriate constant. 

The function F(t) is generally chosen to provide appropriate limiting behaviour for the functional, and parameters are often chosen to fit data, theoretical or experimental.
The functional derivative of this can be expressed most concisely as
\begin{eqnarray}
\lefteqn{ \frac{\delta T^{enh}_s[\rho]}{\delta \rho}  =  
\frac{3}{5}(3\pi^2)^\frac{2}{3} \times} \nonumber \\  
& \left( 
 \frac{5}{3} \rho^\frac{2}{3} F(t)
-\frac{8}{3} \frac{|\nabla \rho|^2}{\rho^2} F'(t)
-2 \nabla . \left(\frac{\nabla \rho}{\rho} F'(t) \right)
\right)
\end{eqnarray}
which differs from the expression obtained by a direct application of the usual formulae\cite{7}.
Severe aliasing problems arise if the standard form is applied, as occurs for the exchange-correlation potential\cite{8}, but this alternative analytic form greatly reduces these numerical difficulties.

The above functional has many deficiencies, the primary one being of only limited non-locality.
Explicitly non-local functional have been investigated in the literature, with the introduction of an analytic form that integrates over the contributions from the charge density at each pair of points in real space\cite{9}.
This can be further generalised by adding a third order term that integrates the contributions from triplets of points in space, and higher order terms.
Unfortunately these are generally extremely computationally expensive to evaluate.
One exception to this is a form proposed by Wang and Teter\cite{10}, Perrot\cite{11} and Smargiassi and Madden\cite{12} where the non-local term is expressed as a convolution integral which can be evaluated efficiently in reciprocal space.
This approximate functional is given by
\begin{eqnarray}
T^{nloc}_{\alpha}[\rho]=
 \frac{3}{5}(3\pi^2)^{\frac{2}{3}} \int \rho^{\frac{5}{3}} d^3 {\mathbf{r}}
-\frac{1}{2} \int \rho^{\frac{1}{2}} \nabla^2  \rho^{\frac{1}{2}} d^3 {\mathbf{r}} \nonumber \\
+ \int \rho^{\alpha}({\mathbf{r}}) w_{\alpha}( {\mathbf{r}} - {\mathbf{r}}') \rho^{\alpha}({\mathbf{r}}') d^3 {\mathbf{r}} d^3 {\mathbf{r}}'
\end{eqnarray}
where the first term is the Thomas Fermi contribution, the second the von Weizacker contribution, and the final term a non-local contribution.
The parameter $\alpha$ is arbitrary, and $w_{\alpha}( {\mathbf{r}} - {\mathbf{r}}')$ is chosen such that the functional has the correct linear response for a homogeneous non-interacting electron gas with the same average charge density as the charge density of interest.
The functional derivative of this approximation is given by
\begin{eqnarray}
\lefteqn{
\frac{ \delta T^{nloc}_{\alpha} }{ \delta \rho } = 
 (3\pi^2)^{\frac{2}{3}} \rho^{\frac{2}{3}} 
-\rho^{-\frac{1}{2}} \nabla^2 \rho^{\frac{1}{2}} } \nonumber \\
& & +2 \alpha \rho^{\alpha-1}( {\mathbf r} )
\int \rho^{\alpha}({\mathbf r}') w_{\alpha}({\mathbf r}-{\mathbf r}') d^3 {\mathbf r}'.
\end{eqnarray}
Our aim is to assess the usefulness of these approximations for carrying out the embedding procedure described in the previous section.

\section{Results}
To investigate this partially frozen density approach we examine the first test case, fcc aluminium with a conventional 4 atom cubic unit cell.
The 3 face-centred atoms are taken to be the substrate system, and this structure is solved for first to provide $\rho_2({\mathbf r})$ and $T_s[\rho_2]$.
A standard plane-wave method\cite{13} is used, with a lattice constant of $a_0=4.05$\AA, a plane-wave cut-off
of $200$ eV, $35 {\mathbf k}$ points in the irreducible wedge, and the Goodwin-Needs-Heine local pseudopotential\cite{14}.
Exchange-correlation is described by the LDA.

Once this substrate is constructed the embedded Kohn-Sham calculation is carried out as a standard plane-wave calculation, but with the trial potential given by Eq. (5) and the total energy given by Eq. (2).
These additional terms describing the substrate, and the calculation only involves the three electrons introduced by the embedded atom at the corner of the unit cell; the 9 substrate electrons are taken into account entirely by their charge density.
Parameters of the calculation are chosen to be the same as for the substrate calculation.
It should be made clear that for the substrate calculation we are solving for the lattice of face centred atoms and their accompanying electrons, but for the embedded calculation we are solving for the \emph{entire} fcc system, but only the electrons associated with the embedded (corner) atom are provided with a Kohn-Sham representation.

\begin{table}
\begin{tabular}{lccc} \hline \hline
Functional              & $E[\rho]/eV$ & $\Delta E/eV$ &   R/\%   \\ \hline
$T_{PW86}$              &-59.68        & -1.35         &  7.2     \\
$T_{\frac{1}{2}}^{nloc}$&-58.41        & -0.08         &  1.2     \\
Kohn-Sham               &-58.33        &     -         &     -    \\ \hline
\end{tabular}
\caption{Total energies per atom, and errors in energies and charge density.}
\end{table}

We discuss results for an enhancement factor approximation (Eq. (6)) with the Perdew and Wang '86 enhancement factor\cite{15} ($T_{PW86}$).
\begin{eqnarray}
F(s)&=& (1+1.296s^2+14s^4+0.2s^6)^{\frac{1}{15}} \nonumber \\
   s&=& \frac{1}{2 (3 \pi^2)^{ \frac{1}{3} } } t^{ \frac{1}{2}},
\end{eqnarray}
and the non-local linear-response corrected functional, Eq. (9) with $\alpha=\frac{1}{2}$ ($T_{\frac{1}{2}}^{nloc}$).
Although calculations have been carried out for a large number of different enhancement factor functionals there was no significant difference in the results and any small differences do not affect our conclusions.
Table 1 shows the errors in the total energies, and the charge density expressed as the mean absolute deviation as a percentage of the mean density. 
It is immediately apparent that the non-local functional provides the superior approximation.

From this we conclude that by applying the non-local functional an accurate total energy and charge density can be obtained by this DFT embedding procedure for an essentially metallic and strongly interacting system.
Semi-local enhancement factor functional do not result in a useful accuracy.

Next we move onto the second test case, a model surface/adsorbate system, the type of system we hope to apply the method to in the future.
This system also places far more demand on the method since the charge densities are considerably more inhomogeneous.
We consider a $\sqrt{2} \times \sqrt{2}$ super-cell with one `adsorbate' per cell, a 3-layer $(100)$ slab and 5 equivalent vacuum layers.
The adsorbate is centred on a four-fold hollow site.
The substrate ($\rho_2({\mathbf r})$) is chosen to be the lower two layers, and embedding calculations were performed with three embedded atoms making up the upper surface layer and the adsorbate.
All calculations were performed with $10 {\mathbf k}$ points in the irreducible wedge.
We chose to examine the potential energy curves produced by varying the adsorbate/surface, and compare these with the equivalent full Kohn-Sham results.
We also compare the embedding results with Kohn-Sham results for a 1-layer $(100)$ slab to discover whether the embedding method can reproduce the interaction between the adsorbate + top layer and lower two layers.

\begin{figure}[h]
\epsfbox{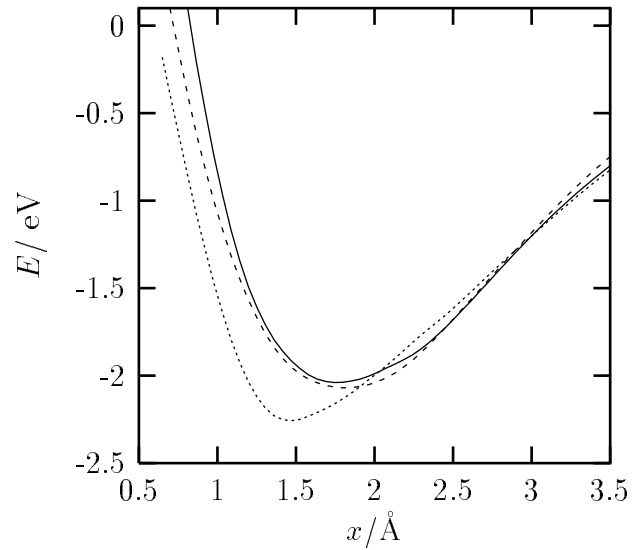}
\caption{Adsorption energy as a function of surface/adsorbate distance vertically above a four-fold hollow.
Dashed line is 3-layer + adsorbate Kohn-Sham result.
Dotted line is 1-layer + adsorbate Kohn-Sham result.
Solid line is results for embedding calculation.
Zero energy is sum of surface and isolated adsorbate total energies.
}
\end{figure}

Figure 1 shows the potential energy curves for the embedded calculation with the non-local functional, for a full Kohn-Sham calculation and for a Kohn-Sham calculation incorporating only one layer of the surface.
Results are promising, with the embedding results agreeing well with the full 3-layer Kohn-Sham calculation.

Equilibrium results for the embedded system are a surface/adsorbate distance of $1.82$ \AA\ and $-2.07$ eV, compared to Kohn-Sham results of $1.77$ \AA\ and $-2.04$ eV - an error of $0.05$ \AA\ and $0.03$ eV respectively.
This differs greatly from the 1 layer/adsorbate system ($1.48$ \AA\ and $-2.26$ eV).
No results are presented for semi-local enhancement factor functionals, since these differ negligibly from the Kohn-Sham 1 layer/adsorbate layer results, indicating that these functionals cannot take into account the coupling between the substrate and adsorbate + top layer for this system.

We conclude that the DFT embedding approach, with a non-local functional, can provide accurate results for two systems with very different charge densities.
Both of these systems are metallic and strongly interacting, so this approach may be useful for investigating a wide range of large systems.

Future work will consist of the application of this method to further systems, both for further validation of the accuracy and the study of large adsorbate/surface systems.

\end{document}